\magnification=1200
\baselineskip=20 pt

\def\l{\Lambda}
\def\dmul{D_{\mu}}
\def\dmuu{D^{\mu}}
\def\dnul{D_{\nu}}
\def\dnuu{D^{\nu}}
\def\piww{\Pi_{ww}^{new}}
\def\pizz{\Pi_{zz}^{new}}
\def\sz{\hat{s}_z}
\def\cz{\hat{c}_z}
\def\lam{\lambda}

\centerline{\bf New Physics, precision electroweak data and an }

\centerline{\bf upper bound on higgs mass}

\centerline{\bf Uma Mahanta}
\centerline{\bf Mehta Research Institute}
\centerline{\bf Chhatnag Road, Jhusi}
\centerline{\bf Allahabad-211019, India}

\centerline{\bf Abstract}

In this paper we express the effect of new physics on gauge boson self
energy corrections through non-renormalizable dimension six operators.
Using the precision electroweak data we then determine a lower bound on the 
scale $\l $ associated with the underlying new physics. The lower bound
on $\l$ is then used to derive an upper bound on $m_h$ through the triviality
relation.

\vfill\eject
\centerline{\bf Introduction}

The Standard Model (SM) of strong and eletroweak interactions based on
the gauge group $SU(3)_c\times SU(2)_l\times U(1)_y$ has been extremely
successful in explaining all the experimental data so far. No significant
deviations from the SM or evidence for new physics has been found so far.
However inspite of this extraordinary phenomenological success many 
theorists consider the SM as incomplete and at best a low energy
description of some underlying high energy theory. The reasons behind 
this attitude are many. The SM does not provide any answer to the hierarchy
problem, to the underlying physics that gives rise to the complex pattern 
of fermion masses or CP violation. In this paper we shall also take 
this attitude and assume that some unknown new physics will manifest itself 
above some high energy scale $\l$ ($\l\gg v$) which will provide answers to
some of these problems. The effect of new heavy phyiscs at scales much
small compared to $\l$ say near the EW scale can be described by
non-renormalizable operators constructed out of the relevant degrees of 
freedom of low energy theory i.e. the SM fields. These non-renormalizable
operators can be systematically constructed by starting from the high
energy theory and integrating out the heavy fields. The effective Lagrangian
for describing physics near the Z pole can therefore be written as [1]

$$L_{eff}=L_{sm}+L_{\l}=L_{sm}+\sum_i c_i {O_i\over \l^{d_i-4}}.\eqno(1)$$

Here $L_{\l}$ represents the effects of new heavy physics at scales 
much smaller than $\l$. $O_i$ is a non-renormalizable operator of
canonical dimension $d_i$. $c_i$ are dimensionless coefficients of
order one if the new physics above $\l$ is strongly coupled which we
assume to be the case. The operator $O_i$ like the renormalizable SM
Lagrangian $L_{sm}$ must be invariant under the 
$SU(3)_c\times SU(2)_l\times U(1)_y$ gauge group and the relevant global
symmetries. In the following we shall assume that the EW gauge symmetry 
is linearly realized on the field content of the theory. 

It is clear that in order to construct non-renormalizable operators
representing the effects of new physics that contribute to the gauge
boson self energy corrections we must concentrate on the gauge-higgs system. 
We shall show that the following dimension six operators contribute to the 
S, T and U parameters.

$$O_1={1\over \l^2} (\dmul \phi )^+{\tau_a\over 2}(\dmuu \phi )
(\phi^+{\tau_a\over 2} \phi ).\eqno(2)$$

$$O_2={1\over \l^2} (\dnul\dmul \phi )^+ (\dnuu\dmuu \phi ).\eqno(3)$$

$$O_3=-{1\over \l^2} W^a_{\mu\nu} B^{\mu\nu} \phi^+{\tau_a\over 2}
\phi .\eqno(4)$$

Note that the negative sign associated with $O_3$ is in accordance
with usual gauge kinetic energy term.

\centerline{\bf S, T and U parameters}

The S, T and U parameters introduced by Peskin and Takeuchi [2]
describe the
effects of new heavy physics, that do not have direct couplings to ordinary
fermions, on the gauge boson self energies. In the $\overline{MS}$ scheme
they are defined as [3,4]

$$\alpha (M_z)T \equiv {\piww (0) \over M_w^2}-{\pizz (0)\over M_z^2}
.\eqno(5)$$

$${\alpha (M_z)\over 4 \sz^2\cz^2} S\equiv  {\pizz (M_z^2)-\pizz (0)\over
M_z^2}.\eqno(6)$$

$${\alpha (M_z)\over 4 \sz^2} (S+U)\equiv  {\piww (M_z^2)-\pizz (0)\over
M_w^2}.\eqno(7)$$

Here $\piww$ and $\pizz$ are the contributions of new heavy physics to
W and  Z self energies. $\sz^2=\sin^2\hat{\theta}_w (M_z)=.2311\pm .0003$ 
in the 
$\overline{MS}$ scheme. Note that T is proportional to the W and Z self 
energies at $q^2=0$ and hence it measures the strength of vector SU(2)
breaking. On the other hand $S(S+U)$ are proportional to the difference 
between Z(W) self energies at $q^2= M_z^2(M_w^2)$ and $q^2=0$. Hence
they measure the strength of axial SU(2) breaking.
The values of S, T and U as determined from the EW data are given by
$$S=-.28\pm .19^{-.08}_{+.17}$$

$$T=-.20\pm .26^{+.17}_{-.12}$$

$$U=-.31\pm .54$$

The first uncertainties are from the inputs. The central values assume 
$m_h=300 $ Gev and the second uncertainty is the change for $m_h=1000 $
GeV (upper) and 60 GeV (lower).

\centerline{\bf Bound on $\l$ from T parameter}

The operator $O_1$ given above contributes to the vector SU(2) breaking
T prameter. In order to see that we write
$$\phi=\left (\matrix{0\cr {v\over \sqrt {2}}\cr}\right ) $$

We then get

$$\eqalignno { O_1&={-v^2\over 8 \l^2}(\dmul\phi )^+{\tau_3\over 2}
(\dmuu\phi ) \phi^+{\tau_3\over 2}\phi  \cr
&={i v^4\over 64 \l^2}(0, 1)[-i g^2
\tau_3 (W_{1\mu}W_1^{\mu}+W_{2\mu}W_2^{\mu}) 
+i (g^2\tau_3 W_{3\mu}W_3^{\mu}\cr
&+2gg^{\prime }W_{3\mu}B^{\mu}+g^{\prime 2}\tau_3 B_{\mu}B^{\mu})]
\left (\matrix {0\cr 1\cr}\right ) \cr
&={v^4\over 64 \l^2}[(g^2+g^{\prime 2})Z_{\mu}Z^{\mu}-2g^2 W^+_{\mu}
W^{-\mu}].&(8)\cr}$$

It follows from the above expression that $\piww (0) =-{\pi\alpha v^4
\over 8\sz^2\l^2}$ and $\pizz (0)= {\pi\alpha v^4\over 8 \sz^2\cz^2 \l^2}$
and hence $T=-{\pi v^4\over 8 \sz^2 \l^2}({1\over M_w^2}+
{1\over \cz^2 M_z^2})$.
Note that due to custodial SU(2) symmetry breaking  by $O_1$, $\piww (0)$
and $\pizz (0)$ turn out to be of opposite sign.
Using the central value of T
as determined from electroweak data we obtain a lower bound of 3.12 TeV
on $\l$.

\centerline{\bf Bound on $\l$ from S parameter}

The operator $O_2$ can contribute to the S parameter. To see that
we replace $\phi$ by its vev. Remembering that for a non-vanishing
contribution to S  we need two powers of $\partial_{\mu}$
we get

$$O_2={1\over 4}[g^2v^2\partial_{\nu} W^{+\mu}
\partial ^{\nu}W^-_{\mu}+{v^2\over 2}
(g^2 +g^{\prime 2} \partial _{\nu}Z^{\mu}\partial ^{\nu}Z_{\mu}+...
].\eqno(9)$$

In the above expression the terms that do not contribute to S 
has been represented by dots.
It follows from the above that $\piww (q^2) = -{1\over 4} 
{g^2v^2\over \l^2} q^2$ , $\pizz (q^2) =-{1\over 4} {g^2+g^{\prime 2}v^2
\over \l^2}q^2$ and hence  $S=-4\pi{v^2\over \l^2}$. Using the central value
of the S  parameter we get $\l\ge $ 1.65 TeV.

Note that the operator $O_2$ does not contribute to the U parameter
which measures the difference between W and Z wavefunction 
renormalizations.

\centerline{\bf Bound on $\l$ from U parameter}

The operator $O_3$ contributes to both S and U. To see that we again
replace $\phi$ by its vev to obtain
$$\eqalignno {O_3 &={v^2\over 4\l^2}W_3^{\mu\nu}B_{\mu\nu}\cr
&={v^2\over 4\l^2}(\sz F_{\mu\nu}+\cz Z_{\mu\nu})(\cz F^{\mu\nu}-\sz
Z^{\mu\nu})+ ...\cr
&=-{v^2\over 2\l^2}\cz\sz \partial_{\mu}Z_{\nu}\partial^{\mu}Z^{\nu}
+...&(10)\cr}$$

Here $F_{\mu\nu}=\partial_{\mu}A_{\nu}-\partial_{\nu}A_{\mu}$
and $Z_{\mu\nu}=\partial_{\mu}Z_{\nu}-\partial_{\nu}Z_{\mu}$.
It follows from the above expression
that $\piww (q^2)=0$ and $\pizz (q^2)=
{v^2\over \l^2}\cz\sz q^2$ and hence $U=-S=- 4{\cz^3\sz^3\over
\alpha (M_z)}{v^2\over \l^2}$. Using the central value of the U
parameter we get a lower bound of 2.75 TeV on $\l$.
Note that the operator $O_3$  also makes a positive contribution to the
S parameter. However since the present data favours a slightly
negative value of S we do not use it to derive a bound on $\l$.
It is interesting to note that the lower bounds on $\l$ determined
from the experimental values of S, T and U are in close agreement
with each other.
\vfill\eject

\centerline{\bf Triviality bound on the higgs mass}

The lower bounds on the $\l$ derived in the previous sections
can be used to derive an upper bound $m_h(\l )$
 on the higgs mass using the 
triviality relation [5]. An approximate perturbative estimate
of  $m_h(\l )$ can be obtained  by using the one loop RG equation 
of the higgs self coupling $\lam $ and assuming that $\lam$ is much 
greater than the other couplings that enter the equation i.e.
$\lam \gg g_t, g, g^{\prime}$. Such an approximation can be justified
for a very heavy higgs boson. In this limit the scalar sector
of the EW theory can be considered in isolation from the gauge bosons 
and the fermions. Let the scalar potential of ths SM be given by

$$V(\phi )=\mu^2 (\phi^+\phi )+\lam (\phi^+ \phi )^2.\eqno(11)$$

where $\mu^2< 0$ and $\lam ={m_h^2\over 2v^2}$ is the higgs self coupling.
The one loop RG equation for $\lam$ in a pure scalar theory is given by

$${d\lam \over dt}={3\lam^2\over 4\pi^2}.\eqno(12)$$

where $t=\ln{Q^2\over Q_0^2}$ and
$Q_0$ is some reference scale which we shall take to be equal to v.
Integrating the above differential equation we get

$${1\over \lam (Q)}= {1\over \lam (Q_0 )}-{3\over 4\pi^2}
\ln{Q^2\over Q_0^2}.\eqno(13)$$

We find that irrespective of how small $\lam (Q_0 )$ is $\lam (Q)$
will ultimately become infinite at some very large energy scale Q.
A bound on $m_h$ can be obtained by requiring that $\lam$ is large but
finite at some large energy scale where new physics appears. This 
consideration leads us to an approximate upper bound on the higgs mass

$$m^2_h<{8\pi^2 v^2\over 3\ln ({\l^2\over v^2})}.\eqno(14)$$.

The perturbative bound on the higgs mass derived from triviality
is cut off dependent in contrast to  the lattice results.
The average value of $\l$ determined from S, T and U parameters
in this paper is equal to 2.5 TeV. The triviality bound on the higgs mass
for this value of $\l$ is 586 GeV. However our perturbativ estimate
will be valid only if the one loop RG equation used by us provides an 
accurate description of the theory at large $\l$. For large $\l$,
howver higher order corrections and non-perturbative effects must be
included to get a more reliable estimate of $m_h(\l )$. Lattice gauge 
theory calculations based on a pure scalar theory gives a limit
$m_h (lattice)< 640 $ GeV [6].

\centerline{\bf References}

\item{1.} W. Buchmuller and D. Wyler, Nucl. Phys. B 268, 621 (1986).

\item{2.} M. Peskin and T. Takeuchi, Phys. Rev. Lett. 65, 964 (1990) and
Phys. Rev. D 46, 381 (1992).

\item{3.} W. Marciano and J. Rosner, Phys. Rev. Lett. 65, 2963 (1990);
D. Kennedy and P. Langacker, Phys. Rev. Lett. 65, 2967 (1990) and
Phys. Rev. D 44, 1591 (1991).

\item{4.} J. Erler and P. Langacker, Phys. Rev. D 52, 441 (1995)
and Phys. Rev. D 54, 103 (1996) Part I.

\item{5.} K. Wilson, Phys. Rev. B 4, 3184 (1971); R. Dashen and
H. Neuberger, Phys. Rev. Lett. 50, 1897 (1983); J. Kuti, L. Lin
and Y. Shen, Phys. Rev. Lett. 61, 678 (1988); M. Luscher and P. 
Weisz, Phys. Lett. B 212, 472 (1988).

\item{6.} A. Hassenfratz, Quantum fields on the computer, Ed.
M. Creutz ( World Scientific, Singapore, 1992) page 125.

\end